\documentclass{elsart}
\usepackage{graphicx}
\usepackage{amsmath}
\usepackage{amsfonts}
\usepackage{amssymb}

\begin{document}
\begin{frontmatter}
\title{On Plasma Oscillations in Strong Electric Fields}
\author[icra]{Remo Ruffini}\ead{ruffini@icra.it},
\author[icra]{Luca Vitagliano}\ead{vitagliano@icra.it},
\author[icra]{She-Sheng Xue}\ead{xue@icra.it}
\address
[icra]{ICRA and
Physics Department, University of Rome ``La Sapienza", 00185 Rome, Italy}
\begin{abstract}
We describe the creation and
evolution of electron-positron pairs in a strong electric field as well as the
pairs annihilation into photons. The formalism is based on generalized Vlasov equations, which are numerically integrated. We recover previous results about the oscillations of the charges,
discuss the electric field screening and the relaxation of the
system to a thermal equilibrium configuration. The timescale of the thermalization is estimated to be $\sim 10^{3}-10^{4}\ \hbar /m_{e}c^{2}$.
\end{abstract}
\end{frontmatter}

Three different earth-bound experiments and one astrophysical observation have
been proposed for identifying the polarization of the electronic vacuum due to
a supercritical electric field ($\mathcal{E}>\mathcal{E}_{\mathrm{c}}\equiv
m_{e}^{2}c^{3}/e\hbar$, where $m_{e}$ and $e$ are the electron mass and
charge) postulated by Sauter-Heisenberg-Euler-Schwinger \cite{S31...}:

\begin{enumerate}
\item  In central collisions of heavy ions near the Coulomb barrier, as first
proposed in \cite{GZ69,GZ70} (see also \cite{PR71,P72,ZP72}). Despite some
apparently encouraging results \cite{S...83}, such efforts have failed so far
due to the small contact time of the colliding ions
\cite{A...95,G...96,L...97,B...95,H...98}. Typically the electromagnetic
energy involved in the collisions of heavy ions with impact parameter
$l_{1}\sim10^{-12}$cm is $E_{1}\sim10^{-6}$erg and the lifetime of the
diatomic system is $t_{1}\sim10^{-22}$s.

\item  In collisions of an electron beam with optical laser pulses: a signal
of positrons above background has been observed in collisions of a 46.6 GeV
electron beam with terawatt pulses of optical laser in an experiment at the
Final Focus Test Beam at SLAC \cite{B...97}; it is not clear if this
experimental result is an evidence for the vacuum polarization phenomenon.
The energy of the laser pulses was $E_{2}\sim10^{7}$erg, concentrated in a
space-time region of spacial linear extension (focal length) $l_{2}\sim
10^{-3}$cm and temporal extension (pulse duration) $t_{2}\sim10^{-12}$s
\cite{B...97}.

\item  At the focus of an X-ray free electron laser (XFEL) (see
\cite{R01,AHRSV01,RSV02} and references therein). Proposals for this
experiment exist at the TESLA collider at DESY and at the LCLS facility at
SLAC \cite{R01}. Typically the electromagnetic energy at the focus of an XFEL
can be $E_{3}\sim10^{6}$erg, concentrated in a space-time region of spacial
linear extension (spot radius) $l_{3}\sim10^{-8}$cm and temporal extension
(coherent spike length) $t_{3}\sim10^{-13}$s \cite{R01}.
\end{enumerate}

and from astrophysics

\begin{enumerate}
\item  around an electromagnetic black hole (EMBH) \cite{DR75,PRX98,PRX02},
giving rise to the observed phenomenon of gamma-ray bursts (GRB)
\cite{RBCFX01a,RBCFX01b,RBCFX01c,RBCFX02}. The electromagnetic energy of an
EMBH of mass $M\sim10M_{\odot}$ and charge $Q\sim0.1M/\sqrt{G}$ is $E_{4}%
\sim10^{54}$erg and it is deposited in a space-time region of spacial linear extension $l_{4}%
\sim10^{8}$cm \cite{PRX98,RV02} and temporal extension (collapse time) $t_{4}\sim10^{-2}$s
\cite{RVX03}.
\end{enumerate}

In addition to their marked quantitative difference in testing the same basic
physical phenomenon, there is a very important conceptual difference among
these processes: the first three occur in a transparency condition in which the
created electron-positron pairs and, possibly, photons freely propagate to
infinity, while the one in the EMBH occurs in an opacity condition
\cite{RSWX00}. Under the opacity condition a thermalization effect occurs and
a final equipartition between the $e^{+}e^{-}$ and $\gamma$ is reached. Far
from being just an academic issue, this process and its characteristic
timescale is of the greatest importance in physics and astrophysics. It has
been shown by a numerical simulation done in Livermore and an analytic work
done in Rome \cite{RSWX00}, that, as soon as the thermalization of $e^{+}%
e^{-}$ and $\gamma$ created around an EMBH has been reached, the plasma self
propells outwards and this process is at the very heart of the gamma-ray burst
(GRB) phenomenon. A critical step was missing up to now: how to bridge the gap
between the creation of pairs in the supercritical field of the EMBH and the
thermalization of the system to a plasma configuration. This letter reports
some progress on this topic with special attention to the timescale needed for
the thermalization of the newly created $e^{+}e^{-}$ pairs in the background
field. The comparison of the thermalization timescale to the one of
gravitational collapse, which occurs on general relativistic timescale, is at
the very ground of the comprehension of GRBs \cite{RVX03}.

The evolution of a system of particle-antiparticle pairs created by the
Schwinger process has been often described by a transport Vlasov equation
(see, for example, \cite{KM85,GKM87}). More recently it has been showed that
such an equation can be derived from quantum field theory
\cite{SRS...97,KME98,SBR...98}. In the homogeneous case, the equations have
been numerically integrated taking into account the back reaction on the
external electric field \cite{KESCM91,KESCM92,CEKMS93,BMP...99}. In many
papers (see \cite{V...01} and references therein) a phenomenological term
describing equilibrating collisions is introduced in the transport equation
which is parameterized by an effective relaxation time $\tau$. In \cite{V...01}
one further step is taken by allowing time variability of $\tau$; the
ignorance on the collision term is then parameterized by a free dimensionless
constant. The introduction of a relaxation time corresponds to the assumption
that the system rapidly evolves towards thermal equilibrium. In this paper we
focus on the evolution of a system of $e^{+}e^{-}$ pairs, explicitly taking
into account the scattering processes $e^{+}e^{-}\rightleftarrows\gamma\gamma
$. Since we are mainly interested in a system in which the electric field
varies on macroscopic length scale ($l\sim10^{8}$cm, above), we
can limit ourselves to a homogeneous electric field. Also, we will use
transport equations for electrons, positrons and photons, with collision
terms, coupled to Maxwell equations. There is no free parameter here: the
collision terms can be exactly computed, since the QED cross sections are
known. Starting from a regime which is far from thermal equilibrium, we find
that collisions do not prevent plasma oscillations in the initial phase of the
evolution and analyse the issue of the timescale of the approach to a
$e^{+}e^{-}\gamma$ plasma equilibrium configuration, which is the most
relevant quantity in the process of gravitational collapse \cite{RVX03}.

The motion of positrons (electrons) is the resultant of three contributions:
the pair creation, the electric acceleration and the annihilation damping. The
homogeneous system consisting of electric field, electrons, positrons and
photons can be described by the equations
\begin{align}
\partial_{t}f_{e}+e\mathbf{E}\partial_{\mathbf{p}}f_{e} &  =\mathcal{S}\left(
\mathbf{E},\mathbf{p}\right)  -\tfrac{1}{\left(  2\pi\right)  ^{5}}%
\epsilon_{\mathbf{p}}^{-1}\mathcal{C}_{e}\left(  t,\mathbf{p}\right)
,\label{pairs}\\
\partial_{t}f_{\gamma} &  =\tfrac{2}{\left(  2\pi\right)  ^{5}}\epsilon
_{\mathbf{k}}^{-1}\mathcal{C}_{\gamma}\left(  t,\mathbf{k}\right)
,\label{photons}\\
\partial_{t}\mathbf{E} &  =-\mathbf{j}_{p}\left(  \mathbf{E}\right)
-\mathbf{j}_{c}\left(  t\right)  ,\label{Maxwell}%
\end{align}
where $f_{e}=f_{e}\left(  t,\mathbf{p}\right)  $ is the distribution function
in the phase-space of positrons (electrons), $f_{\gamma}=f_{\gamma}\left(
t,\mathbf{k}\right)  $ is the distribution function in the phase-space of
photons, $\mathbf{E}$ is the electric field, $\epsilon_{\mathbf{p}}=\left(
\mathbf{p}\cdot\mathbf{p}+m_{e}^{2}\right)  ^{1/2}$ is the energy of an
electron of 3-momentum $\mathbf{p}$ ($m_{e}$ is the mass of the electron) and
$\epsilon_{\mathbf{k}}=\left(  \mathbf{k}\cdot\mathbf{k}\right)  ^{1/2}$ is
the energy of a photon of 3-momentum $\mathbf{k}$. $f_{e}$ and $f_{\gamma}$
are normalized so that $\int\tfrac{d^{3}\mathbf{p}}{\left(  2\pi\right)  ^{3}%
}\ f_{e}\left(  t,\mathbf{p}\right)  =n_{e}\left(  t\right)  $, $\int
\tfrac{d^{3}\mathbf{k}}{\left(  2\pi\right)  ^{3}}\ f_{\gamma}\left(
t,\mathbf{k}\right)  =n_{\gamma}\left(  t\right)  $ , where $n_{e}$ and
$n_{\gamma}$ are number densities of positrons (electrons) and photons,
respectively. The term
\begin{equation}
\mathcal{S}\left(  \mathbf{E},\mathbf{p}\right)  =\left(  2\pi\right)
^{3}\tfrac{dN}{dtd^{3}\mathbf{x}d^{3}\mathbf{p}}=-\left|  e\mathbf{E}\right|
\log\left[  1-\exp\left(  -\tfrac{\pi(m_{e}^{2}+\mathbf{p}_{\perp}^{2}%
)}{\left|  e\mathbf{E}\right|  }\right)  \right]  \delta(p_{\parallel
})\label{S}%
\end{equation}
is the Schwinger source for pair creation (see \cite{KESCM91,KESCM92}):
$p_{\parallel}$ and $\mathbf{p}_{\perp}$ are the components of the 3-momentum
$\mathbf{p}$ parallel and orthogonal to $\mathbf{E}$. We assume that the pairs
are produced at rest in the direction parallel to the electric field
\cite{KESCM91,KESCM92}. We also have, in Eqs. (\ref{pairs}), (\ref{photons}) and
(\ref{Maxwell}),
\begin{align}
\mathcal{C}_{e}\left(  t,\mathbf{p}\right)   &  \simeq\int\tfrac
{d^{3}\mathbf{p}_{1}}{\epsilon_{\mathbf{p}_{1}}}\tfrac{d^{3}\mathbf{k}_{1}%
}{\epsilon_{\mathbf{k}_{1}}}\tfrac{d^{3}\mathbf{k}_{2}}{\epsilon
_{\mathbf{k}_{2}}}\delta^{\left(  4\right)  }\left(  p+p_{1}-k_{1}%
-k_{2}\right)  \nonumber\\
&  \times\left|  \mathcal{M}\right|  ^{2}\left[  f_{e}\left(  \mathbf{p}%
\right)  f_{e}\left(  \mathbf{p}_{1}\right)  -f_{\gamma}\left(  \mathbf{k}%
_{1}\right)  f_{\gamma}\left(  \mathbf{k}_{2}\right)  \right]  ,\label{Ce}\\
\mathcal{C}_{\gamma}\left(  t,\mathbf{k}\right)   &  \simeq\int\tfrac
{d^{3}\mathbf{p}_{1}}{\epsilon_{\mathbf{p}_{1}}}\tfrac{d^{3}\mathbf{p}_{2}%
}{\epsilon_{\mathbf{p}_{2}}}\tfrac{d^{3}\mathbf{k}_{1}}{\epsilon
_{\mathbf{k}_{1}}}\delta^{\left(  4\right)  }\left(  p_{1}+p_{2}%
-k-k_{1}\right)  \nonumber\\
\times &  \left|  \mathcal{M}\right|  ^{2}\left[  f_{e}\left(  \mathbf{p}%
_{1}\right)  f_{e}\left(  \mathbf{p}_{2}\right)  -f_{\gamma}\left(
\mathbf{k}\right)  f_{\gamma}\left(  \mathbf{k}_{1}\right)  \right]
,\label{Cf}%
\end{align}
which describe probability rates for pair creation by photons and pair
annihilation into photons, $\mathcal{M}=\mathcal{M}_{e^{+}\left(
\mathbf{p}_{1}\right)  e^{-}\left(  \mathbf{p}_{2}\right)  \rightleftarrows
\gamma\left(  \mathbf{k}\right)  \gamma\left(  \mathbf{k}_{1}\right)  }$ being
the matrix element for the process $e^{+}\left(  \mathbf{p}_{1}\right)
e^{-}\left(  \mathbf{p}_{2}\right)  \rightarrow\gamma\left(  \mathbf{k}%
\right)  \gamma\left(  \mathbf{k}_{1}\right)  $. Note that the collisional
terms (\ref{Ce}) and (\ref{Cf}) are either unapplicable or negligible in the
case of the above three earth-bound experiments where the
created pairs do not originate a dense plasma. They have been correctly
neglected in previous works (see e. g. \cite{RSV02}). Collisional terms have
also been considered in the different physical context of vacuum polarization by strong chromoelectric
fields. Unlike the present QED case, where expressions for the cross sections
are known exactly, in the QCD case the cross sections are yet unknown and
such collisional terms are of a phenomenological type and usefull uniquely near
the equilibrium regime \cite{V...01}. Finally $\mathbf{j}_{p}\left(
\mathbf{E}\right)  =2\tfrac{\mathbf{E}}{\mathbf{E}^{2}}\int\tfrac
{d^{3}\mathbf{p}}{\left(  2\pi\right)  ^{3}}\epsilon_{\mathbf{p}}%
\mathcal{S}\left(  \mathbf{E},\mathbf{p}\right)  $ and $\mathbf{j}_{c}\left(
t\right)  =2en_{e}\int\tfrac{d^{3}\mathbf{p}}{\left(  2\pi\right)  ^{3}}%
\tfrac{\mathbf{p}}{\epsilon_{\mathbf{p}}}f_{e}\left(  \mathbf{p}\right)  $ are
polarization and conduction current respectively (see \cite{GKM87}). In Eqs.
(\ref{Ce}) and (\ref{Cf}) we neglect, as a first approximation, Pauli blocking
and Bose enhancement (see e.g. \cite{KESCM92}). By suitably integrating
(\ref{pairs}) and (\ref{photons}) over the phase spaces of positrons
(electrons) and photons, we find the following exact equations for mean
values:
\begin{align}
\tfrac{d}{dt}n_{e} &  =S\left(  \mathbf{E}\right)  -n_{e}^{2}\left\langle
\sigma_{1}v^{\prime}\right\rangle _{e}+n_{\gamma}^{2}\left\langle \sigma
_{2}v^{\prime\prime}\right\rangle _{\gamma},\nonumber\\
\tfrac{d}{dt}n_{\gamma} &  =2n_{e}^{2}\left\langle \sigma_{1}v^{\prime
}\right\rangle _{e}-2n_{\gamma}^{2}\left\langle \sigma_{2}v^{\prime\prime
}\right\rangle _{\gamma},\nonumber\\
\tfrac{d}{dt}n_{e}\left\langle \epsilon_{\mathbf{p}}\right\rangle _{e} &
=en_{e}\mathbf{E}\cdot\left\langle \mathbf{v}\right\rangle _{e}+\tfrac{1}%
{2}\mathbf{E\cdot j}_{p}-n_{e}^{2}\left\langle \epsilon_{\mathbf{p}}\sigma
_{1}v^{\prime\prime}\right\rangle _{e}+n_{\gamma}^{2}\left\langle
\epsilon_{\mathbf{k}}\sigma_{2}v^{\prime\prime}\right\rangle _{\gamma
},\nonumber\\
\tfrac{d}{dt}n_{\gamma}\left\langle \epsilon_{\mathbf{k}}\right\rangle
_{\gamma} &  =2n_{e}^{2}\left\langle \epsilon_{\mathbf{p}}\sigma_{1}v^{\prime
}\right\rangle _{e}-2n_{\gamma}^{2}\left\langle \epsilon_{\mathbf{k}}%
\sigma_{2}v^{\prime\prime}\right\rangle _{\gamma},\nonumber\\
\tfrac{d}{dt}n_{e}\left\langle \mathbf{p}\right\rangle _{e} &  =en_{e}%
\mathbf{E}-n_{e}^{2}\left\langle \mathbf{p}\sigma_{1}v^{\prime}\right\rangle
_{e},\nonumber\\
\tfrac{d}{dt}\mathbf{E} &  =-2en_{e}\left\langle \mathbf{v}\right\rangle
_{e}-\mathbf{j}_{p}\left(  \mathbf{E}\right)  ,\label{System1}%
\end{align}
where, for any function of the momenta
\begin{align}
\left\langle F\left(  \mathbf{p}_{1},...,\mathbf{p}_{n}\right)  \right\rangle
_{e} &  \equiv n_{e}^{-n}\int\tfrac{d^{3}\mathbf{p}_{1}}{\left(  2\pi\right)
^{3}}...\tfrac{d^{3}\mathbf{p}_{n}}{\left(  2\pi\right)  ^{3}}\ F\left(
\mathbf{p}_{1},...,\mathbf{p}_{n}\right)  \cdot f_{e}\left(  \mathbf{p}%
_{1}\right)  \cdot...\cdot f_{e}\left(  \mathbf{p}_{n}\right)  ,\\
\left\langle G\left(  \mathbf{k}_{1},...,\mathbf{k}_{l}\right)  \right\rangle
_{\gamma} &  \equiv n_{\gamma}^{-l}\int\tfrac{d^{3}\mathbf{k}_{1}}{\left(
2\pi\right)  ^{3}}...\tfrac{d^{3}\mathbf{k}_{l}}{\left(  2\pi\right)  ^{3}%
}\ G\left(  \mathbf{k}_{1},...,\mathbf{k}_{l}\right)  \cdot f_{\gamma}\left(
\mathbf{k}_{1}\right)  \cdot...\cdot f_{\gamma}\left(  \mathbf{k}_{l}\right)
.
\end{align}
Furthermore $v^{\prime}$ is the relative velocity between electrons and
positrons, $v^{\prime\prime}$ is the relative velocity between photons,
$\sigma_{1}=\sigma_{1}\left(  \epsilon_{\mathbf{p}}^{\mathrm{CoM}}\right)  $
is the total cross section for the process $e^{+}e^{-}\rightarrow\gamma\gamma$
and $\sigma_{2}=\sigma_{2}\left(  \epsilon_{\mathbf{k}}^{\mathrm{CoM}}\right)
$ is the total cross section for the process $\gamma\gamma\rightarrow
e^{+}e^{-}$ (here $\epsilon^{\mathrm{CoM}}$ is the energy of a particle in the
reference frame of the center of mass).

In order to evaluate the mean values in system (\ref{System1}) we need some
further hypotheses on the distribution functions. Let us define $\bar
{p}_{\parallel}$, $\bar{\epsilon}_{\mathbf{p}}$ and $\mathbf{\bar{p}}_{\perp
}^{2}$ such that $\left\langle p_{\parallel}\right\rangle _{e}\equiv\bar
{p}_{\parallel},~\left\langle \epsilon_{\mathbf{p}}\right\rangle _{e}%
\equiv\bar{\epsilon}_{\mathbf{p}}\equiv(\bar{p}_{\parallel}^{2}+\mathbf{\bar
{p}}_{\perp}^{2}+\ m_{e}^{2})^{1/2}$. We assume
\begin{equation}
f_{e}\left(  t,\mathbf{p}\right)  \propto n_{e}\left(  t\right)  \delta\left(
p_{\parallel}-\bar{p}_{\parallel}\right)  \delta\left(  \mathbf{p}_{\perp}%
^{2}-\mathbf{\bar{p}}_{\perp}^{2}\right)  . \label{fe}%
\end{equation}
Since in the scattering $e^{+}e^{-}\rightarrow\gamma\gamma$ the coincidence of
the scattering direction with the incidence direction is statistically
favored, we also assume
\begin{equation}
f_{\gamma}\left(  t,\mathbf{k}\right)  \propto n_{\gamma}\left(  t\right)
\delta\left(  \mathbf{k}_{\perp}^{2}-\mathbf{\bar{k}}_{\perp}^{2}\right)
\left[  \delta\left(  k_{\parallel}-\bar{k}_{\parallel}\right)  +\delta\left(
k_{\parallel}+\bar{k}_{\parallel}\right)  \right]  , \label{fgamma}%
\end{equation}
where $k_{\parallel}$ and $\mathbf{k}_{\perp}$ have analogous meaning as
$p_{\parallel}$ and $\mathbf{p}_{\perp}$ and the terms $\delta\left(
k_{\parallel}-\bar{k}_{\parallel}\right)  $ and $\delta\left(  k_{\parallel
}+\bar{k}_{\parallel}\right)  $ account for the probability of producing,
respectively, forwardly scattered and backwardly scattered photons. Since the
Schwinger source term (\ref{S}) implies that the positrons (electrons) have
initially fixed $p_{\parallel}$, $p_{\parallel}=0$, assumption (\ref{fe})
((\ref{fgamma})) means that the distribution of $p_{\parallel}$ ($k_{\parallel
}$) does not spread too much with time and, analogously, that the distribution
of energies is sufficiently peaked to be describable by a $\delta-$function.
The dependence on the momentum of the distribution functions has been
discussed in \cite{KESCM92,KME98}. Approximations (\ref{fe}), (\ref{fgamma})
reduce Eqs. (\ref{System1}) to a system of ordinary differential equations. In
average, since the inertial reference frame we fix coincides with the center
of mass frame for the processes $e^{+}e^{-}\rightleftarrows\gamma\gamma$,
$\epsilon^{\mathrm{CoM}}\simeq\bar{\epsilon}$ for each species. Substituting
(\ref{fe}) and (\ref{fgamma}) into (\ref{System1}) we find
\begin{align}
\tfrac{d}{dt}n_{e}  &  =S\left(  \mathcal{E}\right)  -2n_{e}^{2}\sigma_{1}%
\rho_{e}^{-1}\left|  {\pi}_{e\parallel}\right|  +2n_{\gamma}^{2}\sigma
_{2},\nonumber\\
\tfrac{d}{dt}n_{\gamma}  &  =4n_{e}^{2}\sigma_{1}\rho_{e}^{-1}\left|  {\pi
}_{e\parallel}\right|  -4n_{\gamma}^{2}\sigma_{2},\nonumber\\
\tfrac{d}{dt}\rho_{e}  &  =en_{e}\mathcal{E}\rho_{e}^{-1}\left|  {\pi
}_{e\parallel}\right|  +\tfrac{1}{2}\mathcal{E}j_{p}-2n_{e}\rho_{e}\sigma
_{1}\rho_{e}^{-1}\left|  {\pi}_{e\parallel}\right|  +2n_{\gamma}\rho_{\gamma
}\sigma_{2},\nonumber\\
\tfrac{d}{dt}\rho_{\gamma}  &  =4n_{e}\rho_{e}\sigma_{1}\rho_{e}^{-1}\left|
{\pi}_{e\parallel}\right|  -4n_{\gamma}\rho_{\gamma}\sigma_{2},\nonumber\\
\tfrac{d}{dt}{\pi}_{e\parallel}  &  =en_{e}\mathcal{E}-2n_{e}{\pi}%
_{e\parallel}\sigma_{1}\rho_{e}^{-1}\left|  {\pi}_{e\parallel}\right|
,\nonumber\\
\tfrac{d}{dt}\mathcal{E}  &  =-2en_{e}\rho_{e}^{-1}\left|  {\pi}_{e\parallel
}\right|  -j_{p}\left(  \mathcal{E}\right)  , \label{System2}%
\end{align}
where $\rho_{e}=n_{e}\bar{\epsilon}_{\mathbf{p}}$, $\rho_{\gamma}=n_{\gamma
}\bar{\epsilon}_{\mathbf{k}}$, ${\pi}_{e\parallel}=n_{e}\bar{p}_{\parallel}$
are the energy density of positrons (electrons), the energy density of photons
and the density of ``parallel momentum'' of positrons (electrons),
$\mathcal{E}$ is the electric field strength and $j_{p}$ the unique component
of $\mathbf{j}_{p}$ parallel to $\mathbf{E}$. $\sigma_{1}$ and $\sigma_{2}$
are evaluated at $\epsilon^{\mathrm{CoM}}=\bar{\epsilon}$ for each species.
Note that Eqs.~(\ref{System2}) are ``classical'' in the sense that the only quantum
information is encoded in the terms describing pair creation and scattering
probabilities. Eqs.~(\ref{System2}) are consistent with energy density
conservation: $\tfrac{d}{dt}\left(  \rho_{e}+\rho_{\gamma}+\tfrac{1}%
{2}\mathcal{E}^{2}\right)  =0.$

\begin{figure}[th]
\begin{center}
\includegraphics[width=8.5cm]{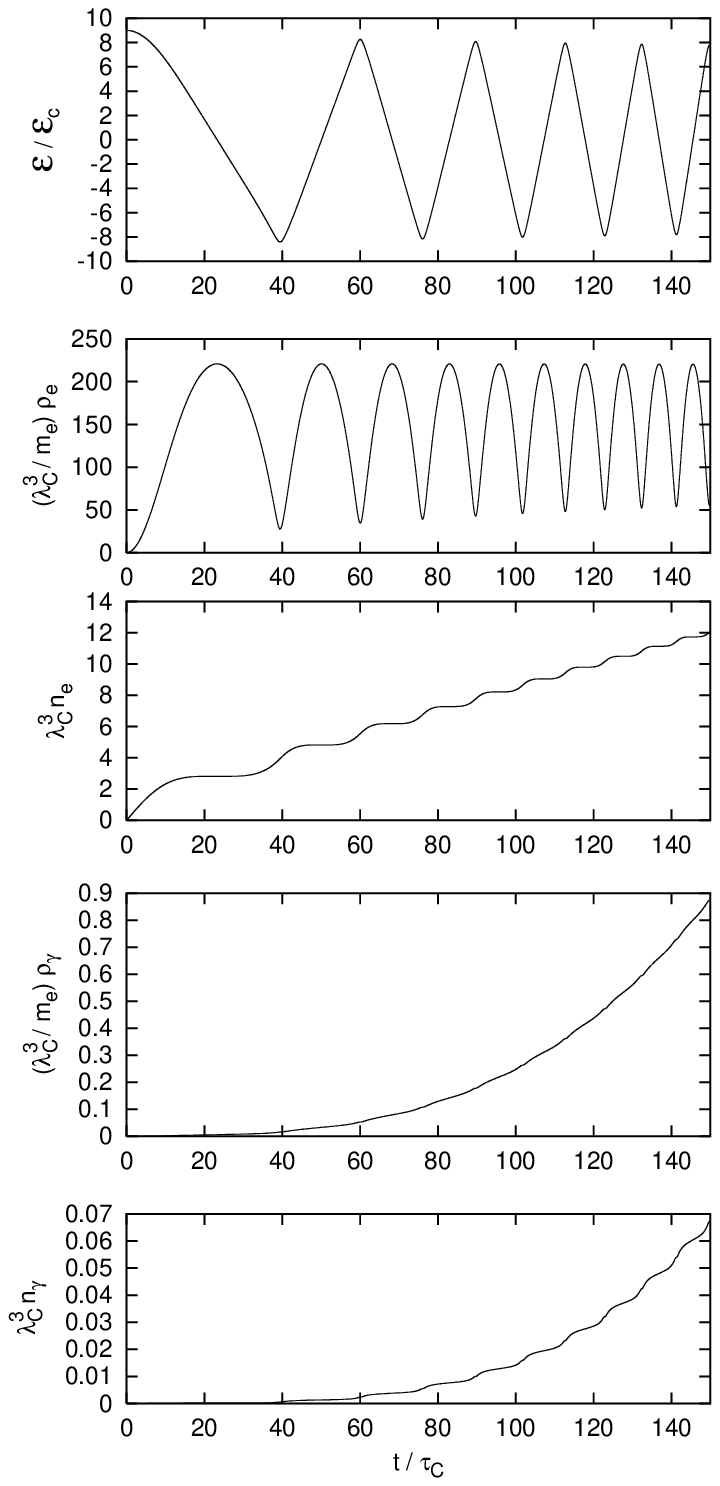}
\end{center}
\caption{Plasma oscillations. We set $\mathcal{E}_{0}=9\mathcal{E}%
_{\mathrm{c}}$, $t<150\tau_{\mathrm{C}}$ and plot: a) electromagnetic field
strength; b) electrons energy density; c) electrons number density; d) photons
energy density; e) photons number density as functions of time.}%
\label{fig1}%
\end{figure}

The initial conditions for Eqs.~(\ref{System2}) are $n_{e}=n_{\gamma}=\rho
_{e}=\rho_{\gamma}=\pi_{e\parallel}=0,~\mathcal{E}=\mathcal{E}_{0}$. In Fig.
\ref{fig1} the results of the numerical integration for $\mathcal{E}%
_{0}=9\mathcal{E}_{\mathrm{c}}$ is showed. The integration stops at
$t=150\ \tau_{\mathrm{C}}$ (where $\tau_{\mathrm{C}}=\hbar/m_{e}c^{2}$). Each
variable is represented in units of $m_{e}$ and
$\lambda_{\mathrm{C}}=\hbar/m_{e}c$. The numerical integration confirms
\cite{KESCM91,KESCM92} that the system undergoes plasma oscillations: a) the
electric field oscillates with decreasing amplitude rather than abruptly
reaching the equilibrium value; b) electrons and positrons oscillates in the
electric field direction, reaching ultrarelativistic velocities; c) the role
of the $e^{+}e^{-}\rightleftarrows$ $\gamma\gamma$ scatterings is marginal in
the early time of the evolution, the electrons are too extremely relativistic
and consequently the density of photons builds up very slowly (see. details in
Fig.~\ref{fig1}).

\begin{figure}[th]
\begin{center}
\includegraphics[width=8.5cm]{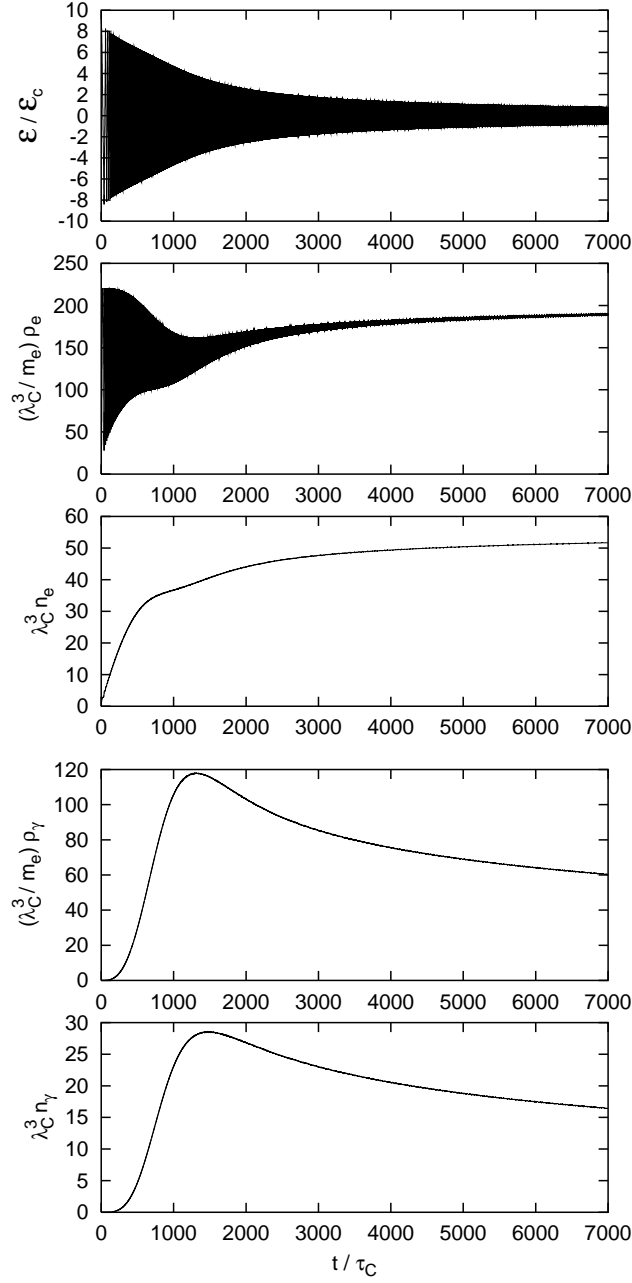}
\end{center}
\caption{Plasma oscillations. We set $\mathcal{E}_{0}=9\mathcal{E}%
_{\mathrm{c}}$, $t<7000\tau_{\mathrm{C}}$ and plot: a) electromagnetic field
strength; b) electrons energy density; c) electrons number density; d) photons
energy density; e) photons number density as functions of time - the
oscillation period is not resolved in these plots. The model used should have
a breakdown at a time much earlier than $7000\tau_{\mathrm{C}}$ and therefore
this plot contains no more than qualitative informations.}%
\label{fig2}%
\end{figure}

At late times the system is expected to relax to a plasma configuration of
thermal equilibrium and assumptions (\ref{fe}) and (\ref{fgamma}) have to be
generalized to take into account quantum spreading of the distribution
functions. It is nevertheless interesting to look at the solutions of
Eqs.~(\ref{System2}) in this regime. In Fig.~\ref{fig2} we plot the numerical
solution of Eqs.~(\ref{System2}) but the integration extends here all the way
up to $t=7000\ \tau_{\mathrm{C}}$ (the time scale of oscillations is not
resolved in these plots). It is interesting that the leading term recovers the
expected asymptotic behaviour: a) the electric field is screened to about the
critical value: $\mathcal{E}\simeq\mathcal{E}_{\mathrm{c}}$ for $t\sim
10^{3}-10^{4}\tau_{\mathrm{C}}\gg\tau_{\mathrm{C}}$; b) the initial
electromagnetic energy density is distributed over electron-positron pairs and
photons, indicating energy equipartition; c) photons and electron-positron
pairs number densities are asymptotically comparable, indicating number
equipartition. At such late times a regime of thermalized
electrons-positrons-photons plasma begins and the system is describable by
hydrodynamic equations \cite{RVX03,RSWX00}.

We provided a very simple formalism apt to describe simultaneously the
creation of electron-positron pairs by a strong electric field $\mathcal{E}%
\gtrsim\mathcal{E}_{c}$ and the pairs annihilation into photons. As discussed
in literature, we find plasma oscillations. In particular the collisions do
not prevent such a feature. This is because the momentum of electrons
(positrons) is very high, therefore the cross section for the process
$e^{+}e^{-}\rightarrow\gamma\gamma$ is small and the annihilation into photons
is negligible in the very first phase of the evolution. As a result, the
system takes some time ($t\sim10^{3}-10^{4}\tau_{\mathrm{C}}$) to thermalize
to a $e^{+}e^{-}\gamma$ plasma equilibrium configuration. We finally remark
that, at least in the case of electromagnetic Schwinger mechanism, the picture
could be quite different from the one previously depicted in literature, where
the system is assumed to thermalize in a very short time (see \cite{V...01}
and references therein).

It is conceivable that in the race to first identify the vacuum polarization
process \emph{\`a la} Sauter-Euler-Heisenberg-Schwinger, the astrophysical observations will reach a positive result before earth-bound experiments, much like in the case of the discovery of lines in the Sun chromosphere by J. N. Lockyer in 1869, later identified with the Helium spectral lines by W. Ramsay in 1895 \cite{G03}.

\quad

We are grateful to A. Ringwald and P. Chen for clarifications on the earth-bound
experiments and general discussions, an anonimous referee for quoting the corresponding analysis in QCD and P. Giannone for the quotation relative to the Helium lines.

\newpage

\newpage

\end{document}